\documentclass[12pt,a4paper]{article}
\pdfoutput=1
\usepackage{jheppub}
\usepackage{amsfonts}
\usepackage{bbm}
\usepackage{graphicx}
\usepackage[utf8]{inputenc}
\usepackage[T1]{fontenc}
\usepackage{amssymb,amsfonts,amsmath}
\usepackage{subfig} 
\usepackage{empheq}
\usepackage{upgreek} 
\usepackage{slashed}

\usepackage{tikz, pgf}
\usetikzlibrary{shapes.misc}
\usetikzlibrary{shapes}

\makeatletter

\pgfdeclareshape{genus}{
 \anchor{center}{\pgfpointorigin}
\backgroundpath{
    \begingroup
    \tikz@mode
    \iftikz@mode@fill

         \pgfpathmoveto{\pgfqpoint{-0.78cm}{-.17cm}}
     \pgfpathcurveto %
            {\pgfpoint{-0.35cm}{-.44cm}}
        {\pgfpoint{0.35cm}{-.44cm}}
        {\pgfpoint{.78cm}{-0.17cm}} 
     \pgfpathmoveto{\pgfqpoint{-0.78cm}{-0.17cm}}
     \pgfpathcurveto %
            {\pgfpoint{-0.25cm}{.25cm}}
        {\pgfpoint{.25cm}{.25cm}}
        {\pgfpoint{0.78cm}{-0.17cm}}
        \pgfusepath{fill}
        \fi

        \iftikz@mode@draw
     \pgfpathmoveto{\pgfqpoint{-1cm}{0cm}}
     \pgfpathcurveto %
            {\pgfpoint{-0.5cm}{-.5cm}}
        {\pgfpoint{0.5cm}{-.5cm}}
        {\pgfpoint{1cm}{0cm}}

     \pgfpathmoveto{\pgfqpoint{-0.75cm}{-0.15cm}}
     \pgfpathcurveto %
            {\pgfpoint{-0.25cm}{.25cm}}
        {\pgfpoint{.25cm}{.25cm}}
        {\pgfpoint{0.75cm}{-0.15cm}}
              \pgfusepath{stroke}
        \fi
        \endgroup
    }
    }

    \makeatother

\font\cmss=cmss10
\font\cmsss=cmss10 at 7pt
\font\manual=manfnt

\newcommand{\bi}{\begin{itemize}}
\newcommand{\ei}{\end{itemize}}

\newcommand{\bea}{\begin{eqnarray}}
\newcommand{\eea}{\end{eqnarray}}
\newcommand{\be}{\begin{equation}}
\newcommand{\ee}{\end{equation}}
\newcommand{\ben}{\begin{eqnarray*}}
\newcommand{\een}{\end{eqnarray*}}
\newcommand{\bem}{\begin{pmatrix}}
\newcommand{\eem}{\end{pmatrix}}
\newcommand{\bl}{\begin{align}}
\newcommand{\el}{\end{align}}
\newcommand{\beg}{\begin{gather}}
\newcommand{\eeg}{\end{gather}}

\newcommand{\cH}{\mathcal{H}}

\newcommand{\IH}{\mathbb{H}}

\newcommand{\TrH[1]}{ {\raise -.5em
                      \hbox{$\buildrel {\textstyle  {\rm Tr } }\over
{\scriptscriptstyle \cH _ {#1}}$}~}}
\newcommand{\TrS}{ {\raise -.5em
                      \hbox{$\buildrel {\textstyle  {\rm Tr } }\over
{\scriptscriptstyle Spec}$}~}}

\newcommand{\res[1]}{{\raise -.5em 
\hbox{$\buildrel{\textstyle{\rm Res}}\over {\scriptscriptstyle {#1}}$}}}
\newcommand{\tends[1]}{{\raise -.5em 
\hbox{$\buildrel{\longrightarrow}\over {\scriptscriptstyle {#1}}$}}}

\def\dbend{\lower3.5pt\hbox{\manual\char127}}

\def\IL{\relax{\rm I\kern-.18em L}}
\def\IH{\relax{\rm I\kern-.18em H}}
\def\rlx{\relax\leavevmode}

\def\ZZ{\rlx\leavevmode\ifmmode\mathchoice{\hbox{\cmss Z\kern-.4em Z}}
 {\hbox{\cmss Z\kern-.4em Z}}{\lower.9pt\hbox{\cmsss Z\kern-.36em Z}}
 {\lower1.2pt\hbox{\cmsss Z\kern-.36em Z}}\else{\cmss Z\kern-.4em
 Z}\fi}

\title{ Ramanujan and Quantum Black Holes}

\author[1, 2, 3]{Atish Dabholkar,}

\affiliation[1]{International Centre for Theoretical Physics\\
Strada Costiera 11, Trieste 34151 Italy
\vspace{4pt}}

\affiliation[2]{  Sorbonne Université, CNRS, \\
Laboratoire de Physique Théorique et Hautes Énergies, \\LPTHE, F-75005 Paris, France
\vspace{4pt}}

\affiliation[3]{CNRS, UMR 7589, LPTHE,  Paris, F-75005 France
\vspace{4pt}}

\abstract{
Explorations of quantum black holes in string theory have led to fascinating connections with the work of  Ramanujan on partitions and mock theta functions, which in turn relate to diverse  topics in number theory and enumerative geometry.   This article aims to explain the physical  significance of these interconnections.   \\
\leftline{}
\leftline{}
\leftline{}
\centerline{ \textit{ Contribution to `Encyclopedia of Srinivasa Ramanujan and His Mathematics'}}
\\
\centerline{\emph{Based on  `Nag Memorial Lectures' delivered at the Institute for Mathematical Sciences}}
\\
}
\keywords{black holes, partitions, mock modular forms, string theory}

\begin{document}
\maketitle

\section{Quantum Black Holes}

A classical black hole  is  the region of spacetime  which cannot send signals to  faraway observers.  It is black because  even light  cannot escape its strong gravity. To make these notions precise, consider the prototypical Schwarzschild spacetime with  line element \cite{Wald:1984rg}
\begin{eqnarray} \label{metric}
ds^{2 } =  - (1-\frac{2M}{r}) dt^{2} + (1-\frac{2M}{r})^{-1}dr^{2} +  r^{2} (d\theta^{2} + \sin^{2} \theta d\phi^{2}) \, ,
\end{eqnarray}
where $t$ is time and $r, \theta, \phi$ are spherical coordinates. This pseudo-Riemannian metric  is Ricci flat and hence satisfies   Einstein equations without matter.  The  region $r \leq 2M$  is the Schwarzschild black hole of mass $M$. The boundary of this region  at $r=2M$  is  called the \textit{event horizon}. It is  a peculiar 3-surface that is  both stationary (independent of time) and lightlike  (its conormal $dr$   has vanishing norm). 
 Light emanating from  inside  the black hole cannot  escape to faraway region at large $r$ because it cannot overtake the event horizon moving at the speed of light. 

One consequence of this unusual causal structure is  the discovery of Bekenstein and Hawking \cite{Bekenstein:1973ur, Hawking:1974sw} that a  black hole carries entropy $S(M)$ given by  
\begin{equation}\label{entropy}
S(M) =\frac{A(M)}{4} \, .
\end{equation} 
This remarkable formula,  valid in the limit of large area,  implies a surprising connection between thermodynamics, geometry, and quantum mechanics.  
It poses two important questions. 
\begin{enumerate}
\item In quantum theory,  a system  with entropy $S(M)$ and mass $M$ corresponds to an ensemble of vectors  in an eigensubspace $\mathcal{H}(M)$  of dimension $d(M)$ with  mass eigenvalue $M$.  Can one associate such a subspace with a black hole in the Hilbert space of quantum gravity and compute its dimension? 
\item Can one define  and compute \cite{Wald:1993nt,Sen:2008vm} a quantum generalization of  (\ref{entropy}) that is valid even for small area? Schematically, it is expected to have the form
\begin{eqnarray}\label{quantum}
S = a_{0 } A  + a_{1} \log (A) + \frac{a_{2}}{A} + \ldots   b_{0}e^{-c_{0} A} + \ldots
\end{eqnarray} 
with some coefficients $( a_{0}, b_{0}, c_{0}\ldots)$.
\end{enumerate}
Both questions have rich implications and have provided  invaluable clues in the search for quantum gravity.

String theory offers the most promising approach to  a consistent quantum theory of gravity. Equations of motion of string theory are generalizations of Einstein equations.  Some of the  `supersymmetric' solutions of these equations are ten-dimensional product manifolds $\mathcal{M}_{10} = \mathcal{M}_{4}(Q) \times \Sigma_{6}$ where $\Sigma_{6}$ is a compact Ricci-flat six-dimensional manifold. The manifold $\mathcal{M}_{4}(Q)$ contains a black hole with  a metric analogous to  (\ref{metric}) but now specified by a vector of  integral charges $Q$ so that the mass is determined by $Q$.  The  entropy $S(Q)$ of  this special class of   black holes is  determined entirely in terms of the charges and geometric properties of $ \mathcal{M}_{4}(Q)$ by a formula analogous to (\ref{quantum}). 

The fundamental physical significance of  entropy  of these black holes stems from the Boltzmann relation
\begin{equation}\label{Boltzmann}
d(Q) =  \exp[{S(Q)}] \, 
\end{equation} 
which  links a   \textit{macroscopic}  geometric property of spacetime to the underlying \textit{microscopic}  Hilbert space of quantum gravity.  It  provides a window into  the quantum  structure of spacetime  in much the same way  entropy of  gases revealed  the quantum structure of matter. 

Within the framework of string theory, a black hole in $ \mathcal{M}_{4}(Q)$ corresponds to a multi-dimensional `membrane' wrapping a homology cycle in $\Sigma_{6}$. The Hilbert space $\mathcal{H}(Q)$ is the space of states of this membrane. The integers $d(Q)$, sometimes called the \textit{degeneracy}, are then given by certain 
enumerative invariants of $\Sigma_{6}$,
 which in turn are related to problems in combinatorics and number theory. 
This line of enquiry leads naturally to the work of Ramanujan as we illustrate below. 

\section{Colored Partitions}

A simple example is $\Sigma_{6}= K3\times T^{2}$ where $K3$ is a Ricci-flat 4-manifold (`Kummer surface') and $T^{2}$ is a 2-torus.
The integer $d(Q)$ in this case equals the Euler character of the symmetric product of $n$ copies of $K3$, where   $n$ is a particular integral norm of the vector $Q$. 
Recall that the Euler character $\chi_{1}$ of a single copy $K3$ is 24 and $K3$ has only even cohomology, that is, it admits only even harmonic forms.  Given this topological data, the problem of computing the Euler character $\chi_{n}$ of $Sym^{n} (K_{3})$ is equivalent \cite{Dabholkar:1989jt,Vafa:1994tf} to the problem of finding the number of  partitions $p_{24}(n)$ of a positive integer $n$ using integers of $24$ different colors, a problem  close to Ramanujan's work. The solution is given \cite{Goettsche:1990go} in terms of $q$-coefficients of a partition function:
\begin{equation}
Z(\tau) = \frac{1}{\Delta(\tau)}  \,,  \quad (q:= e^{2\pi i \tau})
\end{equation}
where $\Delta(\tau)$ is the Ramanujan cusp form of weight $12$. The partition function is thus a modular form of weight $-12$:
\begin{equation}
Z(\frac{a \tau +b}{c\tau +d}) = (c\tau +d )^{-12} Z (\tau) \,  
\end{equation}
for all 
\begin{equation}\label{modular}
\left(
       \begin{array}{cc} a & b \\ c & d \\  
       \end{array} \right) \in SL(2,\textbf{Z})\,, 
\end{equation}
and admits a  Fourier expansion
\begin{equation}
Z (\tau )=\sum_{n=0}^{\infty} C(n)q^{n-1} \, .
\end{equation}
It is  easy combinatorics to see that
\begin{equation}
p_{24}(n) =  C(n) \quad  (n > 0) \, .
\end{equation}

\section{Hardy-Ramanujan Formula}

The modular properties of  $Z(\tau)$  imply that $C(n)$ admits the  Hardy-Ramanujan-Rademacher expansion \cite{Rademacher:1964ra}  and equals
\begin{equation}\label{kloos1}
\sum_{c=1}^{\infty} \left(\frac{2\pi}{c}\right)^{14}K(n, -1, c) \,  {I}_{13}\left(\frac{z}{c}\right) \, 
\end{equation} 
with $z= 4\pi\sqrt{n}$, where
\begin{equation}\label{intrep}
 I_{13}(z) :=\frac{1}{2\pi i}\int_{\epsilon-i\infty}^{\epsilon+i\infty} \, \frac{dt}{t^{14}}\exp [{t +\frac{z^2}{4t}}]
\, \nonumber
\end{equation}
is a modified Bessel function and 
\begin{equation}\label{Kloosterman}
 K(n, m, c) :=\sum_{\substack{ d \in \textbf{Z} /c \textbf{Z}\\
                  d a =1\,\text{mod}\,c}}e^{2\pi i \left(n \frac{d}{c}+ m \frac{a}{c}\right)} \nonumber 
\end{equation}
is the  Kloosterman sum for $n, m, c \in \textbf{Z}$. 
%

For a large class of other black holes, the integers $d(Q)$ are similarly related \cite{Sen:2007qy} to more complicated invariants  such as the Donaldson-Thomas invariants. They are given in terms of Fourier coefficients of a variety of modular objects such as Jacobi  or Siegel forms, and  admit an expansion \cite{Moore:2004fg, Manschot:2007ha} that generalizes (\ref{Kloosterman}). 

Remarkably, the Hardy-Ramanujan formula is  an \textit{exact} convergent expansion of an integer in terms of analytic functions. This is just what is needed to verify the Boltzmann relation (\ref{Boltzmann}) where $d(Q)$ is an integer given by a counting problem but $S(Q)$  is  an analytic function determined by the geometry of spacetime. 
Its leading asymtoptics valid for large charges has exponential growth. For example, the large $z$ asymptotics gives 
\be
d(n) \sim I(z) \sim e^{z} \sim e^{4\pi \sqrt{n}} \, , \qquad n \gg 1
\ee
This is in accordance with leading, large area   entropy of black holes and the Boltzmann relation. 
Using methods of supersymmetric localization it has become possible to compute both sides of equation (\ref{Boltzmann}) in a number of examples \cite{Sen:1995in,Strominger:1996sh, Maldacena:1999bp,Mandal:2010cj, Dabholkar:2011ec, Dabholkar:2014ema} to find nontrivial agreement. 
This  verification of Boltzmann relation even beyond the large area approximation constitutes one of the important successes of string theory.

\section{Mock Jacobi Forms}

In an interesting class of  examples, the partition function $Z(\tau, z)$ depends  on an additional  variable  $z$. It  is modular in $\tau$ 
 and transforms like a  \textit{meromorphic} Jacobi form of index $m$ under the `elliptic'transformations 
\begin{equation}
z \rightarrow z + \lambda \tau + \mu  \, , \quad \lambda, \mu \in \textbf{Z} \, ,
\end{equation}
 with double poles at $z=0$ and its images. The Fourier coefficients $d(Q)$ are no longer uniquely defined but depend on the choice of the $z$-contour. 
This seems to contradict (\ref{Boltzmann}) because entropy of $S(Q)$ of the  corresponding  black hole is uniquely defined and does not suffer from any ambiguities. 


%

The resolution of this puzzle has to do with the existence of \textit{multi-centered} black holes. Equations of string theory admit a  two-centered black hole solution that depends on the moduli of $\Sigma_{6}$.  Locally, each center looks like a single black hole but the two centers are bound together.   The distance between them is fixed by the charges and  the moduli, and diverges as one approaches a co-dimension one  `wall'  in the moduli space. The solution no longer exists on the other side of the wall. 
For a given charge $Q$, the enumerative `invariants' $d(Q, \mu)$ now have a mild dependence on the moduli.  They are invariant in a given chamber in the moduli space  but jump upon crossing a wall separating two chambers because on one side the counting includes the two-centered black holes but on the other side it does not.   This is known as the \textit{wall-crossing} phenomenon \cite{Denef:2007vg,Kontsevich:2008fj}. 

The ambiguity in defining the Fourier coefficients of $Z(\tau, z)$ now has a nice physical interpretation \cite{Sen:2007vb, Dabholkar:2007vk, Cheng:2008fc}. 
The choice of the Fourier contour depends on the moduli $\mu$ of  $\Sigma_{6}$ in a specific way.  Crossing a wall in the moduli space  corresponds to crossing a pole of  the Fourier integrand.  The residue at the pole gives the difference in the enumerative invariants  $d(Q, \mu)$ on the two sides of the wall.

There is a special \textit{attractive} chamber in the moduli space  which admits only single-centered black holes as solutions. Let us denote the moduli in this chamber by $\mu^{*}$.  One can now pose a more refined question whether one can compute  $d(Q, \mu^{*})$ and if it satisfies  (\ref{Boltzmann}). The answer to this question  naturally leads us into the realm of mock modular forms \cite{Zwegers:2002} and is provided by  the following theorem \cite{Dabholkar:2012nd}. 
The partition function admits a unique decomposition
\begin{equation}\label{decomposition}
Z(\tau, z)=Z^{F}(\tau, z) +Z^{P}(\tau, z)\,,
\end{equation}
such that the polar part has  the form
\begin{equation}\label{Appel}
Z^{\rm P}(\tau, z) \; = \; \frac{p_{24}(m+1) }{\Delta(\tau)} \,  A_{m}(\tau, z)
\end{equation}
where 
\begin{equation}\label{Appel}
A_{m}(\tau, z) = \sum_{s\in\textbf{Z}} \, \frac{q^{ms^2 +s}y^{2ms+1}}{(1 -q^s  y)^{2}}\end{equation}
is called an Appell-Lerch sum. 
It admits a \textit{completion} obtained a correction term that is \textit{nonholomorphic} in $\tau$ :
\begin{equation}\label{Appel2}
\hat A(\tau, z) = A(\tau, z) + A^{*}_{m}(\tau, z)
\end{equation}
which transforms as a  Jacobi form. The completion satisfies
\begin{eqnarray}\label{anomaly}
& &\sqrt{\frac{8 \pi i}{m}}  \tau_2^{3/2} \, \frac{\partial} {\partial \bar{\tau}}   \, \hat A_{m}(\tau, z)  \\
&\qquad & =  
 - \sum_{{\ell\,\textrm{mod}\, 2m}}  {\overline{\vartheta_{m,\ell}(\tau)}} \,\, \vartheta_{m,\ell} (\tau,z) \, . \nonumber
\end{eqnarray}
It also implies that $Z^{F}(\tau, z)$ by itself is  not modular but  admits a modular completion $\hat Z^{F}(\tau, z)$ which  transforms like a  Jacobi form  and satisfies a `holomorphic anomaly equation' similar to (\ref{anomaly}). Such an object is called a (mixed) mock Jacobi form. It turns out that Ramanujan's  mock theta functions are closely related to another type of mock Jacobi forms very similar to the ones that appear in the context of quantum black holes. 

Using these ingredients one  obtains a beautifully consistent picture \cite{Dabholkar:2012nd}.  The degeneracy  $d(Q, \mu*)$ of single-centered black holes is given by the Fourier coefficients of the \textit{mock} Jacobi form
which are independent of the moduli. They are uniquely defined by the charges and satisfy the Boltzmann relation (\ref{Boltzmann}). The degeneracy of multi-centered black holes is given by the Fourier coefficients of $Z^{\rm P}(\tau, z)$. These  do depend on the choice of the contour and hence indirectly on the moduli, and  jump upon crossing walls in the moduli space consistent with the wall-crossing phenomenon. 

It is remarkable that the mathematical ideas and tools created by Ramanujan a century ago  have now come to have  deep  applications in quantum gravity. 



\bibliographystyle{JHEP}
\bibliography{ramanujan}
\end{document}